\documentclass[doublecol]{epl2}
\newcommand{\be}{\begin{equation}}\newcommand{\ee}{\end{equation}}
\newcommand{\ba}{\begin{array}{l}}\newcommand{\ea}{\end{array}}
\newcommand{\baa}{\begin{eqnarray}}\newcommand{\eaa}{\end{eqnarray}}
\newcommand{\re}[1]{(\ref{#1})}
\newcommand{\ci}[1]{\cite{#1}}

\newcommand{\sn}{sn}
\newcommand{\dn}{dn}
\newcommand{\cn}{cn}
\newcommand{\am}{am}

\title{Branched Josephson junctions:\\ Current carrying solitons in  external magnetic fields}
\shorttitle{Branched Josephson junctions} 

\author{D. Matrasulov\inst{1,2} \and K. Sabirov\inst{3} \and D. Babajanov\inst{1} \and H. Susanto\inst{4}}
\shortauthor{D. Matrasulov\etal}

\institute{
  \inst{1} Turin Polytechnic University in Tashkent - 17 Niyazov Str., 100095,  Tashkent, Uzbekistan\\
  \inst{2} Center for Nanotechnology - National University of Uzbekistan, 100174, Tashkent, Uzbekistan\\
  \inst{3} Tashkent University of  Information Technology - Amir Temur Avenue 108, Tashkent 100200, Uzbekistan\\
  \inst{4} Department of Mathematical Sciences, University of Essex - Wivenhoe Park, Colchester CO4 3SQ, UK.

} 

\abstract{ We consider branched Josephson junction created by
planar superconductors connected to each other through the
Y-junction insulator.  Assuming that the structure interacts with
the external constant magnetic field, we study static sine-Gordon
solitons in  such system by modeling them in terms of the
stationary sine-Gordon equation on metric graph. Exact analytical
solutions of the problem are obtained and their stability is
analyzed.}

\begin{document}

\maketitle

\section{Introduction}

Low dimensional nanoscale materials are the basic structures for
many electronic devices. Optimization of their electronic
properties and effective functioning of such devices require
tuning the material properties and revealing most appropriate
device architecture. This concerns also superconducting structures
such as  Josephson junctions. Remarkable feature of Josephson
junctions is the fact that the phase difference at the junction is
described in terms of the sine-Gordon equation (see, e.g.
\ci{Jos1}-\ci{Imanbey}). This makes them powerful testing ground
for experimental realization of sine-Gordon solitons
\ci{ablowitz1}-\ci{Kivshar1}. So far, different models have been
proposed for the study of static and traveling solitons using
Josephson junctions \ci{Malomed1} -\ci{Ustinov2}.

In this paper we address the problem of static solitons in
branched Josephson junction containing planar superconductors
connected to each other via the branched insulators having the
shape of Y-junction. The system is considered as interacting with
constant external magnetic field. The phase differences on each
branch of such structure is described in terms of the stationary
sine-Gordon equation on metric graphs. Earlier, in the
Ref.\ci{Karim2018} we considered a version of such system for the
case of absence of  current carrying states. Unlike to that case,
in the present study, including current leads to completely
different vertex boundary conditions, and hence, to different
solutions than those obtained in \ci{Karim2018}.   Provided
certain constraints given in terms of the system parameters, we
obtain exact analytical solutions of the stationary sine-Gordon
equation on metric graphs, modeling static solitons in branched
Josephson junction. Motivation for the study of such model comes
from several practically important problems, such as
superconducting quantum interference devices (SQUID in networks),
superconducting qubits in networks, as well as granular
superconductors. Among others, most attractive practical
application could be experimental realization of sine-Gordon
solitons in networks. We note that the soliton dynamics in
networks is becoming one of the hot topics in nonlinear and
mathematical physics \ci{Hadi1,Hadi2,Zarif}- \ci{tbcnlse}.
Refs.\ci{Hadi1,Hadi2} considered for the first time the
sine-Gordon equation on branched domain for modeling Josephson
junction at tricrystal surfaces. Integrable sine-Gordon equation
on metric graphs is studied in \ci{caputo14,Our1,Karim2018}.
Linear and nonlinear systems of PDE on metric graphs  are
considered in \ci{Bolte,KarimBdG,KarimNLDE}.\\
Among different realizations of Josephson junctions those having
the discrete and branched structure is of special importance, as
it allows to study soliton dynamics in discrete systems and
networks. The early treatment of superconductor networks
consisting of Josephson junctions meeting at one point dates back
to \ci{Nak1}. An interesting realization of Josephson junction
networks at tricrystal boundaries  was discussed earlier in
\ci{Kogan}, which inspired later detailed study of the problem
using the sine-Gordon equation on networks in \ci{Hadi1,Hadi2}.
Some versions of Josephson junction networks containing chain of
the linear superconductors connected via the point-like
insulators, have been studied on the basis of discrete sine-Gordon
model \ci{Luca1}-\ci{Ovchinnik}. Unlike the previously discussed
versions of Josephson junction networks, our model is simple from
the viewpoint of experimental realization and can be studied .\\
 The paper is organized as follows. In the next section we give a formulation of the problem in terms of
the sine-Gordon equation on metric graphs. Section III presents
the derivation of exact analytical solutions for special cases and
their stability analysis.  Finally, Section IV presents some
concluding remarks.

\begin{figure}[ht!]
\includegraphics[width=80mm]{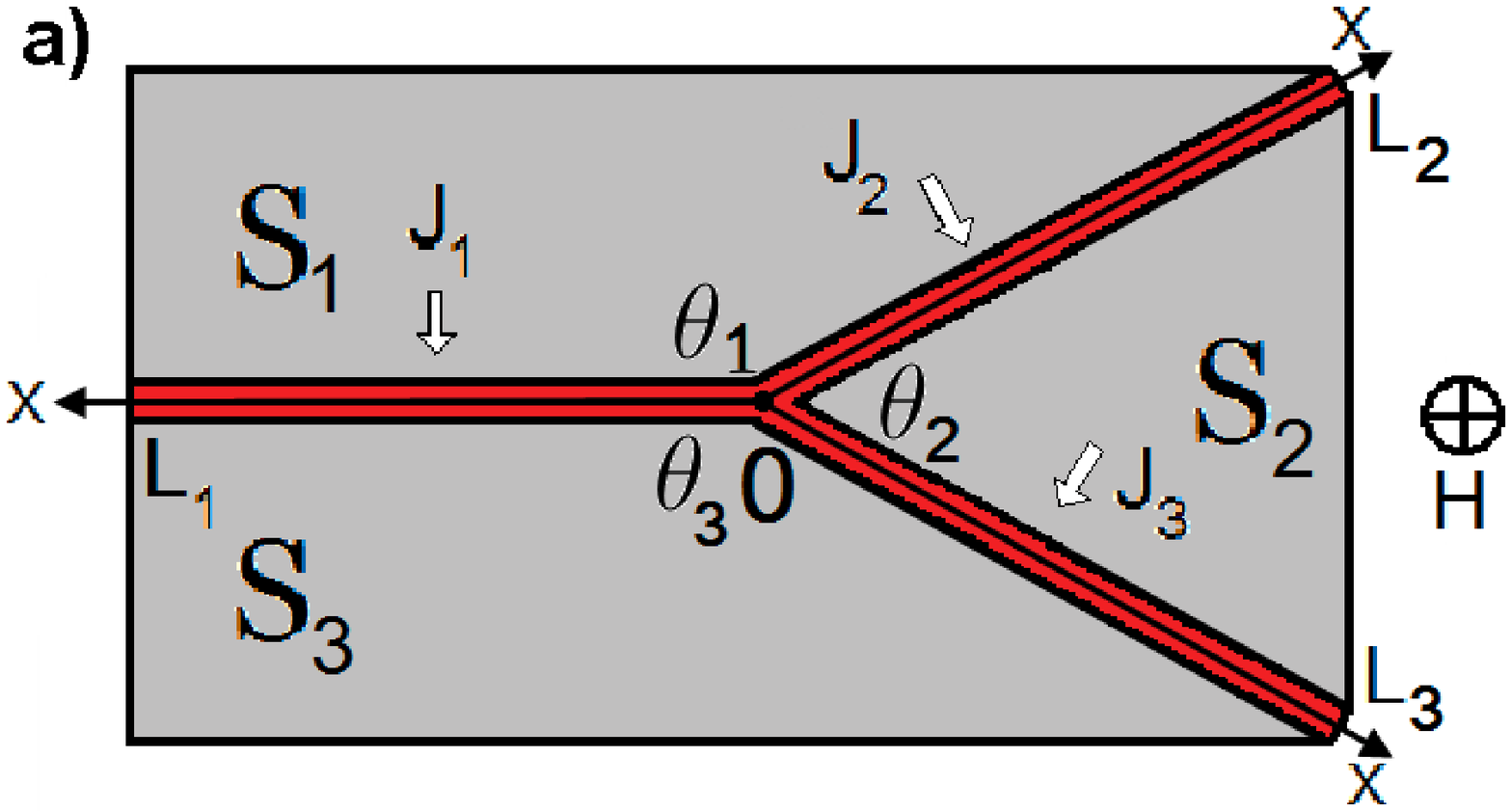}
\includegraphics[width=80mm]{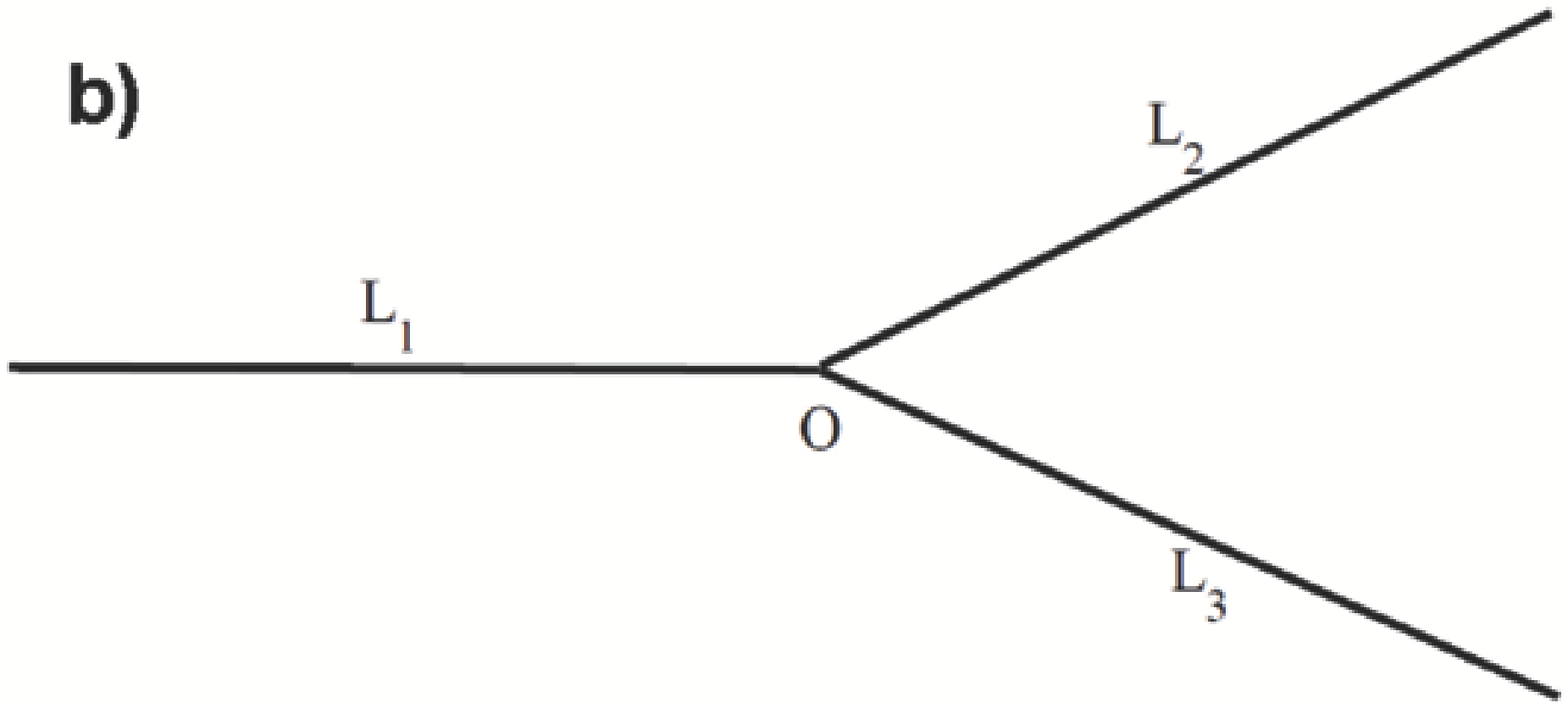}
\caption{(Color online) (a) Branched Josephson Y-junction in a
constant magnetic field, $H$. Red lines imply normal metal or
insulator. $J_1, J_2$ and $J_3$ are the Josephson currents flowing
through each branch of the junction. b) Basic star graph. $L_j$ is
the length of the $j$th branch of the graph($j=1,2,3$).}
\label{pic1}
\end{figure}

\section{Modeling of branched Josephson junction in terms of metric
graph}

Consider the structure presented in Fig. 1a, which represents a
Josephson junction consisting of three  planar superconductors
connected to each other via the branched insulator in the form of
Y-junction. The whole system is assumed to interact with external
constant magnetic field, $H$ which is perpendicular to the plane
of superconductors. Such structure can be considered as the
branched version of the Josephson junction considered in the
Refs.\ci{Kupl06,Kupl07}. The structure can be modeled in terms of
metric star graph having three branches, i.e., simple Y-junction(
see, Fig. 1b). For each bond of the star graph a coordinate $x_j$
is assigned. The origin of coordinates at the vertex, 0 and for
bonds we put $x_j\in[0;L_j]$. Then on can use shorthand notation
$\phi_j(x)$ for $\phi_j (x_j)$, where $x$ is the coordinate on the
bond $j$ to which the component $\phi_j$ refers. The phase
difference on each branch  $\phi_j$, is described in terms of the
stationary sine-Gordon equation on metric star graph
\ci{Karim2018}: \be
\frac{d^2}{dx^2}\phi_j=\frac{1}{\lambda_j^2}\sin(\phi_j),\,0<x<L_j,
\label{ssge1} \ee where $j=1,2,3$ is the bond (branch) number and
the origin of coordinates is assumed at the branching point, $O$.
To solve this equation, one needs to impose boundary conditions at
the branching point, $O$. Such boundary conditions can be derived
from the physical properties of the structure presented in Fig.
1a. Computing, at the branching point, the phase differences,
$\phi_1=\theta_1-\theta_3$, $\phi_2=\theta_1-\theta_2$,
$\phi_3=\theta_2-\theta_3$, where $\theta_{1,2,3}$ are the phases
on each superconductor, one can obtain first set of the vertex
boundary conditions given by

\begin{equation}
\phi_1|_{x=0}-\phi_2|_{x=0}-\phi_3|_{x=0}=0.\label{ssgebc002}
\end{equation}

In the following we will use  the system of units $\hbar=c=2\pi
d=e=1$, where $d$ is equal to twice the penetration depth (for
identical superconductors) plus the insulator (or normal metal)
thickness \ci{Owen}. In such units, e.g., for $d=1mm$ $J_j=1$ is
equal to $\approx 7.64 nA$, and for the magnetic field $H=1$
implies that $H \approx 1.22 \mu A/m$, etc.

Then the local magnetic field in terms of $\phi_j$ can be written
as \be h_j(x)=\frac{\partial \phi_j}{\partial x},\ee where we have
scaled the local magnetic field over $\pi$ (i.e.
$\frac{h_j(x)}{\pi}\to h_j(x)$). The current density on each
branch of the junction is given as \ci{Kupl07,Owen,Zharkov}
\begin{equation}
j_j(x)=\frac{1}{4\lambda_j^2}\sin\phi_j(x).\label{cd1}
\end{equation}
Integrating Eq. (\ref{cd1}) over the each bond and using Eq.
(\ref{ssge1}) we can find the current on each bond as \ci{Owen}
\begin{eqnarray}
J_j=\frac{1}{4}\left(\left.\frac{d\phi_j}{dx}\right|_{x=L_j}-\left.\frac{d\phi_j}{dx}\right|_{x=0}\right).\label{cd2}
\end{eqnarray}

Using continuity of the local magnetic field $h_j(x)$ at the
branching point ($h_1(0)=h_2(0)=h_3(0)$) we get the second set of
vertex boundary conditions: \be
\left.\frac{d\phi_1}{dx}\right|_{x=0}=\left.\frac{d\phi_2}{dx}\right|_{x=0}=\left.\frac{d\phi_3}{dx}\right|_{x=0}.\label{ssgebc001}
\ee

For complete formulation of the problem, one needs also to impose
boundary conditions at the end of each branch. This can be done by
writing explicitly the value of local magnetic field in terms of
external and intrinsic magnetic field. These latter are supposed
to be induced by Josephson current on each branch. Denoting this
magnetic field on each branch by $H_{Jj}$ ($j=1,2,3$) we have the
following Neumann type boundary conditions at the end of each
branch:

\begin{eqnarray}
\left.\frac{d\phi_1}{dx}\right|_{x=L_1}=H+H_{J1},\nonumber\\
\left.\frac{d\phi_2}{dx}\right|_{x=L_2}=H-H_{J2},\nonumber\\
\left.\frac{d\phi_3}{dx}\right|_{x=L_3}=H-H_{J3}.
\label{ssgebc003}
\end{eqnarray}

Writing the same expression at the branching point, one can derive
explicit relation expressing  the external magnetic field, $H$ in
terms of the derivatives of phase differences:
\begin{eqnarray}
H=\frac{1}{4}\sum_{j=1}^3\left.\frac{d\phi_j}{dx}\right|_{x=L_j}+\frac{1}{4}\left.\frac{d\phi_1}{dx}\right|_{x=0}.\label{mf1}
\end{eqnarray}

The problem given by Eqs.\re{ssge1}, \re{ssgebc002},
\re{ssgebc001} and \re{ssgebc003} completely determines  the
problem of sine-Gordon equation on metric star graph, which is the
model  for the static solitons in branched Josephson junction
presented in Fig.1a.

Exact solutions of Eq.\re{ssge1}  for the boundary conditions
providing the absence of current-carrying states ($J_j=0$), have
been obtained in \ci{Karim2018}, where the stability of such
solutions also was analyzed.  Here we consider current carrying
states ($J_j\neq 0$) in the branched Josephson junction, which are
described by different boundary conditions.

\section{Static solitons and their stability}

The problem given by Eqs. \re{ssge1}, \re{ssgebc002},
\re{ssgebc001} and \re{ssgebc003} have different types of
solutions. However, only the stable solutions of this problem can
be considered as the physical ones. These latter describe the
phase difference in branched Josephson junction in  Fig.1a.
Therefore, following the Refs.\ci{Kupl06,Kupl07}, we provide
prescription for stability analysis for the solutions of
Eq.\re{ssge1}.  Starting point for such analysis is the Gibbs
free-energy functional which can be written as \ci{Kupl06,Kupl07}
\begin{equation}
\Omega_G={\sum}_{j=1}^{3}\Omega_G^{(j)}\left[\phi_j,\frac{d\phi_j}{dx};H,H_{J1},H_{J2},H_{J3}\right],\label{gfe1}
\end{equation}
where $\Omega_G^{(j)}$ is the Gibbs free energy functional on each
bond (see the Ref.\ci{Kupl04} for details of the derivation of
$\Omega_G$), which is given by
\begin{eqnarray}
\Omega_G^{(j)}\left[\phi_j,\frac{d\phi_j}{dx};H,H_{J1},H_{J2},H_{J3}\right]=2H^2L_j-\nonumber\\
-\left(H\pm H_{Jj}\right)\phi_j(L_j)+\nonumber\\
+\left(H-H_{J1}+H_{J2}+H_{J3}\right)\phi_j(0)+\nonumber\\
+\int\limits_0^{L_j}\left[\frac{1}{\lambda_j^2}\left(1-\cos\phi_j(x)\right)+\frac{1}{2}\left(\frac{d\phi_j(x)}{dx}\right)^2\right]dx,\label{gfe2}
\end{eqnarray}
where we take the "+" sign for $j=1$, and "-" sign for other
cases.
 Eq.\re{ssge1} together with the boundary conditions \re{ssgebc002}, \re{ssgebc001}, \re{ssgebc003}  follows from the condition
\begin{equation}
\delta\Omega_G=0 .\label{fv1}
\end{equation}

Criterion for the stability of the solution of problem given by
Eqs.\re{ssge1}, \re{ssgebc002}, \re{ssgebc001} and \re{ssgebc003},
can be obtained from the second variation of $\Omega_G,$ i.e.,
from $$\Omega_G=0,$$ which leads to the following Sturm-Liouville
problem \ci{Kupl06,Kupl07,Karim2018}:
\begin{eqnarray}
-\frac{d^2\psi_j}{dx^2}+\frac{1}{\lambda_j^2}\cos\phi_j(x)\psi_j=\mu\psi_j,\,0<x<L_j,\nonumber\\
\psi_1|_{x=0}-\psi_2|_{x=0}-\psi_3|_{x=0}=0,\nonumber\\
\left.\frac{d\psi_1}{dx}\right|_{x=0}=\left.\frac{d\psi_2}{dx}\right|_{x=0}=\left.\frac{d\psi_3}{dx}\right|_{x=0},\nonumber\\
\left.\frac{d\psi_j}{dx}\right|_{x=L_j}=0,\,j=1,2,3,\label{slp1}
\end{eqnarray}
where $\psi_j =\delta\phi_j,\;\; j=1,2,3.$ In terms of the lowest
eigenvalue, $\mu_0$,  the criterion for stability of the solution
can be formulated as follows. If $\mu_0 < 0,$ the solution
$\phi_j(x)$ corresponds to a saddle point of Eq.\re{gfe1} which
implies that the solution is absolutely unstable and unphysical.
Stable (physical) solutions correspond to the case, when $\mu_0 >
0,$ ($\delta^2\Omega_G> 0$). The boundaries of the stability
regions for these solutions is determined by the condition $\mu_0
= 0$ ($\delta^2\Omega_G = 0$), that leads to the following
Sturm-Liouville problem:
\begin{eqnarray}
-\frac{d^2\bar\psi_j}{dx^2}+\frac{1}{\lambda_j^2}\cos\phi_j(x)\bar\psi_j=0,\,0<x<L_j,\label{slp2}\\
\bar\psi_1|_{x=0}-\bar\psi_2|_{x=0}-\bar\psi_3|_{x=0}=0,\label{slp3}\\
\left.\frac{d\bar\psi_1}{dx}\right|_{x=0}=\left.\frac{d\bar\psi_2}{dx}\right|_{x=0}=\left.\frac{d\bar\psi_3}{dx}\right|_{x=0},\label{slp4}\\
\left.\frac{d\bar\psi_j}{dx}\right|_{x=L_j}=0,\,j=1,2,3.\label{slp5}
\end{eqnarray}
Using Eqs.\re{slp2}-\re{slp5}, one can  explicitly find the
stability boundary for each type of solution of the problem given
by Eqs. \re{ssge1}, \re{ssgebc002}, \re{ssgebc001} and
\re{ssgebc003}.

General solution of Eq.\re{ssge1} can be obtained from the
following first integral \ci{Kupl06,Kupl07}: \be
\frac{1}{2}\left[\frac{d\phi_j}{dx}\right]^2 +\cos\phi_j =C_j,\;\;
-1 \leq C_j < \infty, \ee with $C_j$ being the integration
constant. Depending on the value of $C_j$ this general solution
can be determined as type I and II. Namely, for $C_j\in [-1,1)$ we
have solution of type I, while solution of type II corresponds to
the values, $C_j\in [1, \infty)$. Both solutions for $H\neq 0,$
and $J_j=0$ have been found in \ci{Karim2018} where it was shown
that only the special case of the solution of type II is stable.
Following the Refs. \ci{Kupl06,Kupl07}, instead of $C_j$ we
introduce new parametrization constant, $k_j$, which is defined,
for the solution of type I as
$$
k_j^2 \equiv \frac{1+C_j}{2}, \; -1< k_j< 1,
$$
and
$$
k_j^2 \equiv \frac{2}{1+C_j}, \; -1 < k_j < 1,
$$
for solution of type II. General (type I) solution  of
Eq.\re{ssge1} can be written as \ci{Kupl06,Kupl07,Karim2018}
\begin{equation}
\phi_j(x)=(2n_j+1)\pi+2\arcsin\left\{k_j\sn\left[\frac{x-x_{0j}}{\lambda_j},k_j\right]\right\}\label{gensol1}
\end{equation}
 where $\sn$ is Jacobi  elliptic function \ci{ASbook}, and $x_{0j}$ are integration constants
which obey the constraints given by the following inequality:
$$
-\lambda_jK(k_j) < x_{0j} < \lambda_jK(k_j),\;\;j=1,2,3.
$$
Solution given by  Eq. (\ref{gensol1}) fulfils  the vertex
boundary conditions given by Eqs. (\ref{ssgebc002}),
(\ref{ssgebc001}) and (\ref{ssgebc003}), i.e., becomes exact
analytical solution of the problem given by Eqs. (\ref{ssge1}),
(\ref{ssgebc002}), (\ref{ssgebc001}) and (\ref{ssgebc003}),
provided the following constraints hold true:
\begin{eqnarray}
\lambda_1=\lambda_2=\lambda_3=\lambda,\,k_1=k_2=k_3=k,\nonumber\\
-\frac{x_{01}}{\lambda_1}=\frac{x_{02}}{\lambda_2}=\frac{x_{03}}{\lambda_3}=x_0.\label{eq6}
\end{eqnarray}

\begin{equation}
n_1=n_2+n_3.\label{numbers}
\end{equation}
 The solution \re{gensol1}  can be stable only for those values
of $k$ which belong to the interval $[k_c, 1)$ ($k_c\sn
[x_0,k_c]=\frac{1}{2}$). Therefore in the following, in analogy
with that in the Ref.\ci{Kupl07}, we compute the physical
characteristics of the system at $k = k_c(x_0)$ which correspond
to its values at the stability border. Using the relation \be
\frac{d\phi_j(x)}{dx}=-\frac{2k}{\lambda}\cn\left[\frac{x}{\lambda}\pm
x_{0},k\right],\label{eq5} \ee and Eq. (\ref{ssgebc003}), when
$j=1$ we take $+$ sign , when $j=2,3$ we take $-$ sign, the
stability border for the current-carrying states can be written
as:
\begin{eqnarray}
J_j^{(c)}=-\frac{k_c}{2\lambda}\left(\cn\left[\frac{L_j}{\lambda}\pm x_0,k_c\right]-\cn\left[x_0,k_c\right]\right),\label{cd4}\\
H^{(c)}=-\frac{k_c}{2\lambda}\left(\sum_{j=1}^3\cn\left[\frac{L_j}{\lambda}\pm
x_0,k_c\right]+\cn\left[x_0,k_c\right]\right).\label{mf2}
\end{eqnarray}

\begin{figure}[ht!]
\includegraphics[width=85mm]{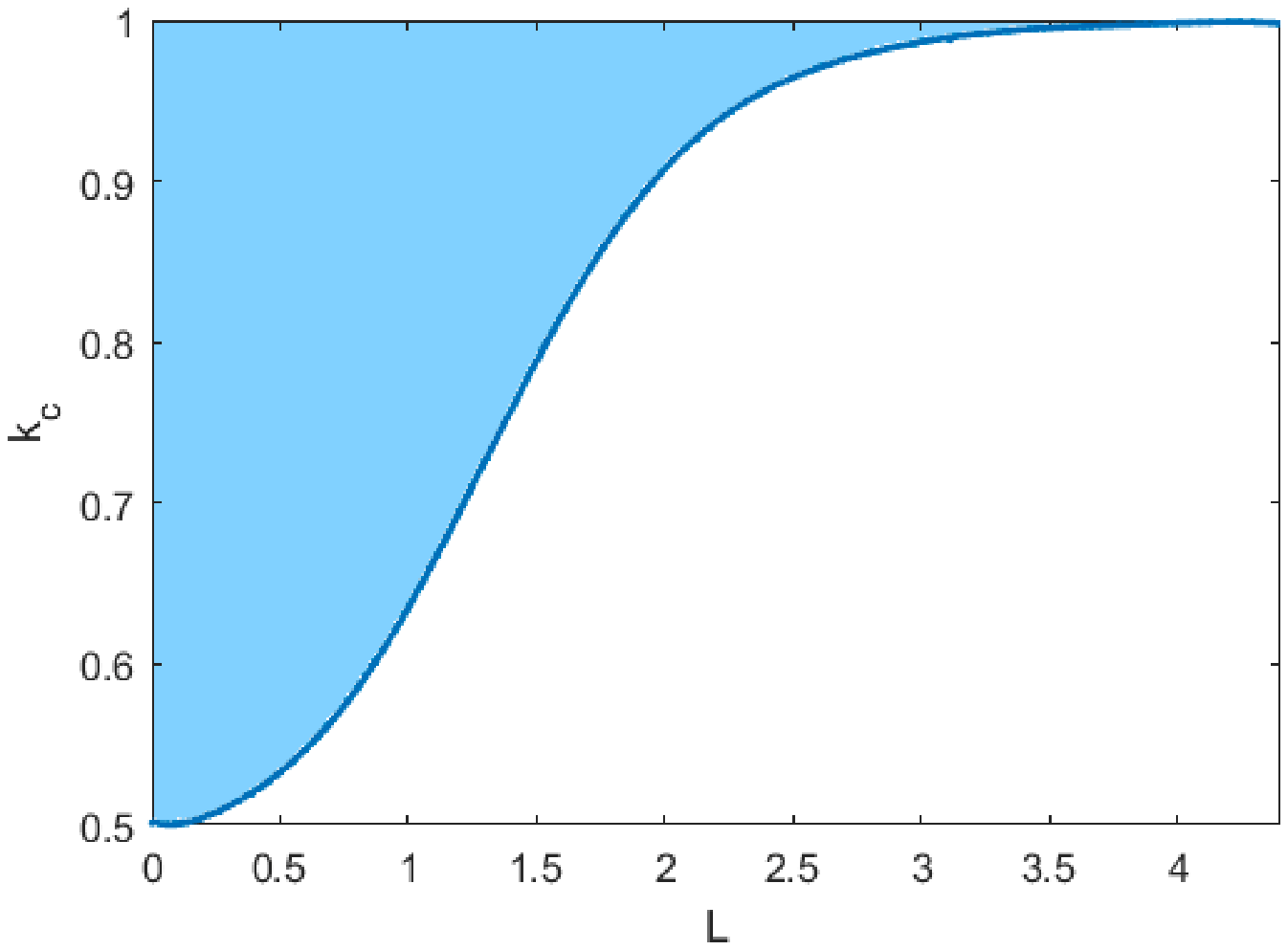}
\includegraphics[width=70mm]{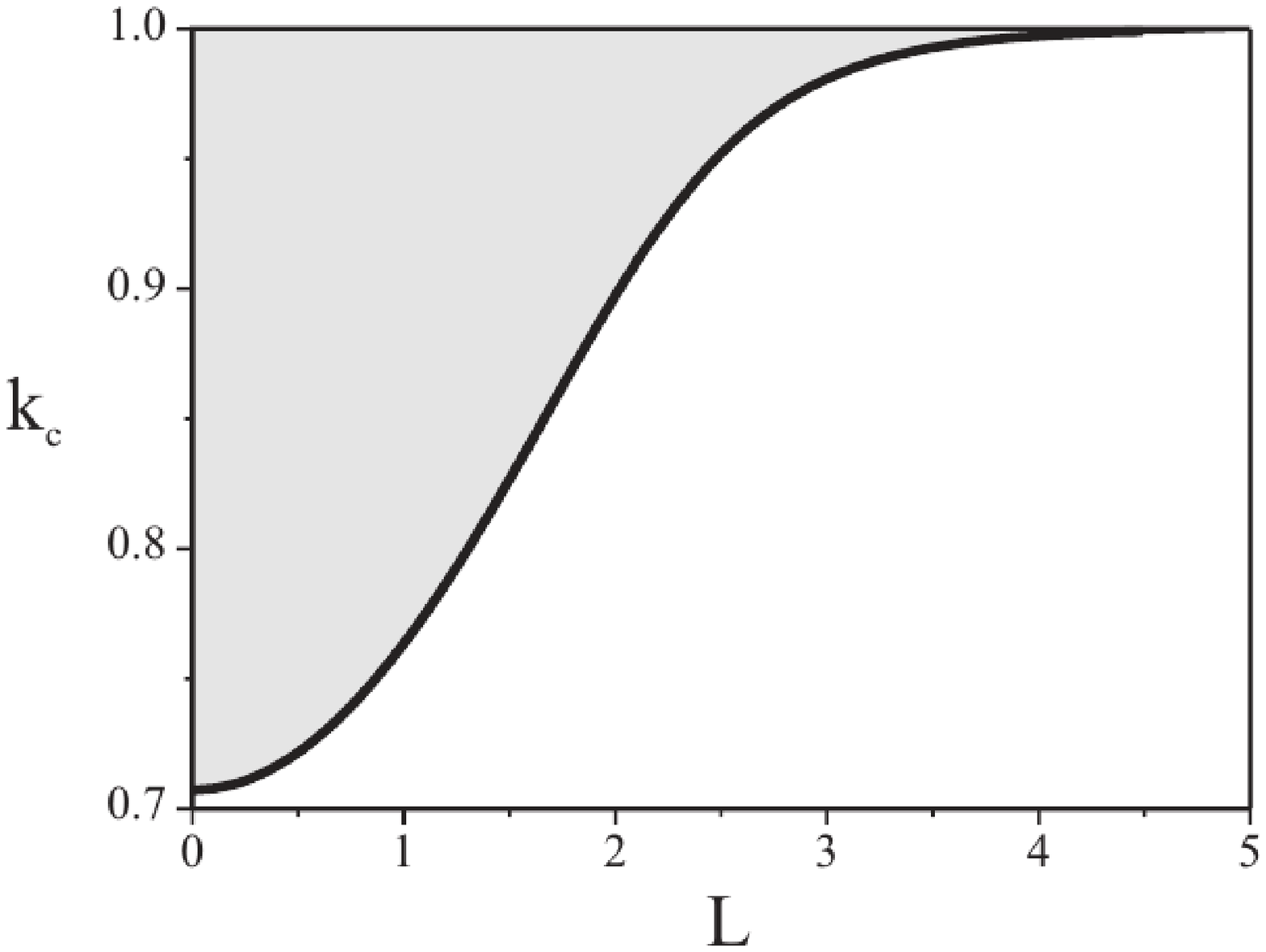}
\caption{(Color online) Upper panel: The dependence of the
stability border, $k_c=k_c(L)$  on the branch length,(solid line)
for the branched Josephson junction. Colored area corresponds to
the stability area. Lower panel: similar plot in the linear
(unbranched) case from the Ref.\ci{Kupl07}.}
 \label{pic4}
\end{figure}

\begin{figure}[ht!]
\includegraphics[width=85mm]{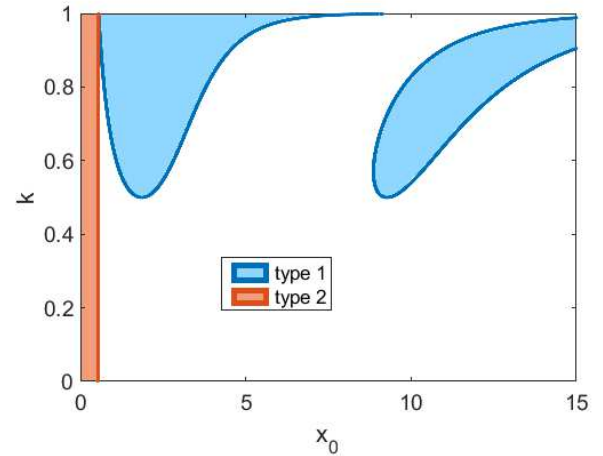}
\caption{(Color online) The stability region (colored) of $\phi$
in the parametric plane. Branch lengthes are $L_1=1, L_2=2,
L_3=3$.}\label{pic2}
\end{figure}

Fig.\ref{pic4} presents plot of  $k_c$  as a function of the
parameter, $L$ determined from $L_1=L, L_2=2L, L_3=3L$. The left
(colored) area of each plot corresponds to the stability region.
Lower panel in this figure presents corresponding plot for linear
case from the Ref.\ci{Kupl07}.  Since $k_c$ appears as the value
of $k$ at which the Sturm-Liouville (stability) problem has zero
($\mu_0= 0$) eigenvalue,  it is important to check, at which
values of $x_0$ this is possible. Fig. \ref{pic2} presents plot of
$k_c$ as a function of $x_0$, i.e.,  the stability region of
$\phi$ in the parametric plane. Colored area corresponds to the
stability region.\\ The solution of type II can be treated
similarly to that of type I, by considering two cases. The case
$H>0,\,J_j =0$ has been studied in detail in the Ref.
\ci{Karim2018}. Therefore we drop this part. Here we will focus on
the case $H>0,\,J_j>0$. General (type II) solution for this case
can be written as \begin{equation}
\phi_j(x)=\pi(2n_j+1)+2\am\left(\frac{x-x_{0j}}{\lambda_jk_j},k_j\right).\label{stsol2}
\end{equation}
Fulfilling the boundary conditions given by Eqs. (\ref{ssgebc002})
and (\ref{ssgebc001}) leads to the constraints in Eqs.\re{eq6} and
\re{numbers}.  Stable solutions and the border between stability
and unstable regions can be determined similarly to that for
solution type I.

From Eqs. (\ref{cd2}) and (\ref{mf1}) we get the expressions for
current and magnetic field
\begin{eqnarray}
J_j=\frac{1}{2\lambda k}\left(\dn\left[\frac{L_j}{\lambda k}\pm x_0,k\right]-\dn\left[x_0,k\right]\right),\label{cd5}\\
H=\frac{1}{2\lambda k}\left(\sum_{j=1}^3\dn\left[\frac{L_j}{\lambda k}\pm x_0,k\right]+\dn\left[x_0,k\right]\right),\label{mf3}\\
x_0\in[0;x_{0,c}].\label{eq12}
\end{eqnarray}

\begin{figure}[ht!]
\includegraphics[width=60mm,height=130mm]{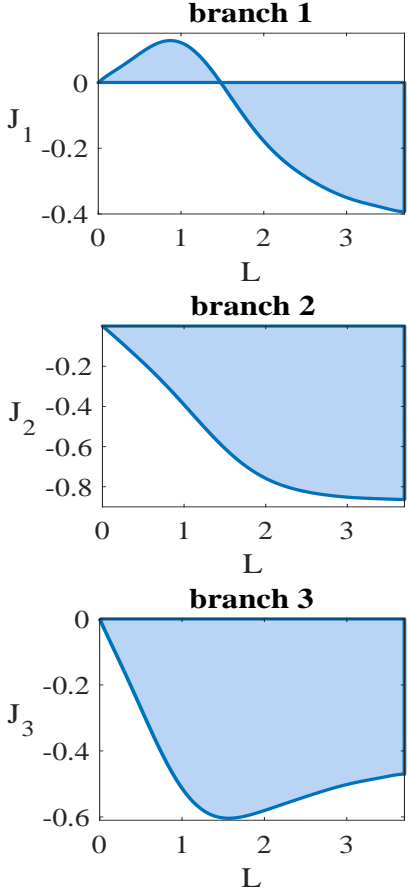}
\caption{(Color online) The dependence $J_c=J_c(L)$ for $H=0$
(solid line). The stability region is colored, parameters are the
same as in Fig.\ref{pic4}.} \label{pic5}
\end{figure}

\begin{figure}[ht!]
\includegraphics[width=75mm]{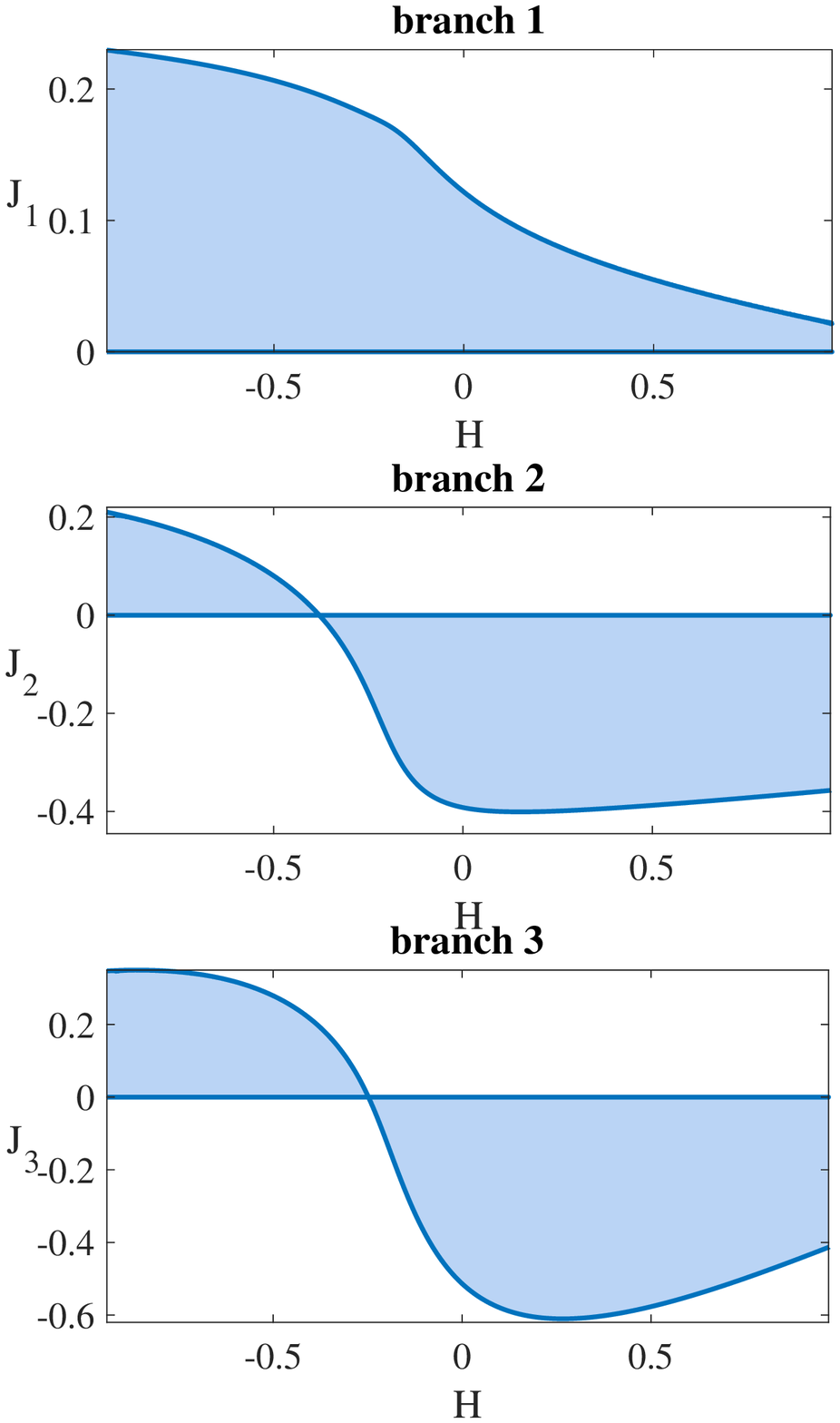}
\caption{(Color online) The stability region (colored) of $\phi_j$
in the physical plane $(J,H)$ for type 1 solutions and the same
parameters as in Fig.3.} \label{pic6}
\end{figure}

\begin{figure}[ht!]
\includegraphics[width=80mm, height=120mm]{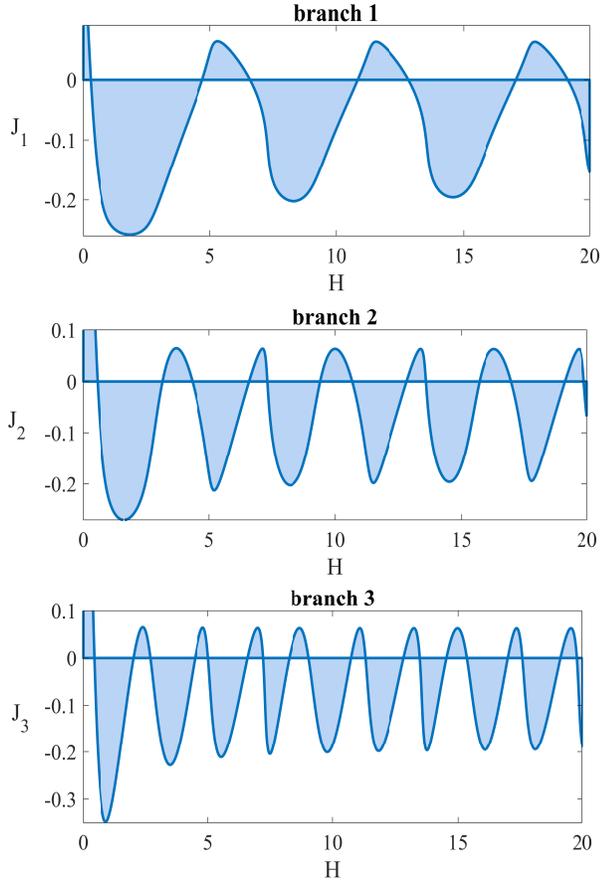}
\caption{(Color online) The stability region (colored) of $\phi_j$
in the physical plane $(J,H)$ for type 2 solutions and the same
parameters as in Fig.3.} \label{pic7}
\end{figure}

In Fig. \ref{pic5}, the dependence of the current on the branch
length, $L_j$ is plotted. Colored (lower) parts  corresponds to
the the stability area. Figs. \ref{pic6} and \ref{pic7} present
the plots of the current, $J_j$ as a function of the magnetic
field for type 1 and type 2, respectively. The colored area in
each plot corresponds to the stability region, i.e, presents the
stability region of $\phi_j$ in the physical plane $(J,H)$.

\begin{figure}[ht!]
\includegraphics[width=80mm]{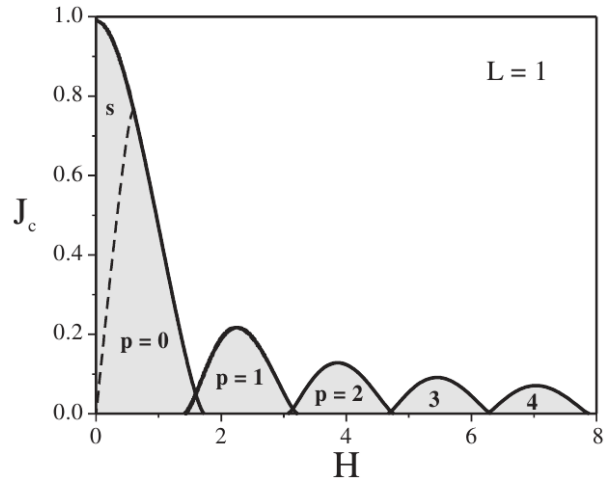}
\caption{ The stability region of $\phi_j$ in the physical plane
$(J,H)$ for linear (unbranched) Josephson junction from
Ref.\ci{Kupl07}.} \label{pic8}
\end{figure}

\begin{figure}[ht!]
\includegraphics[width=70mm]{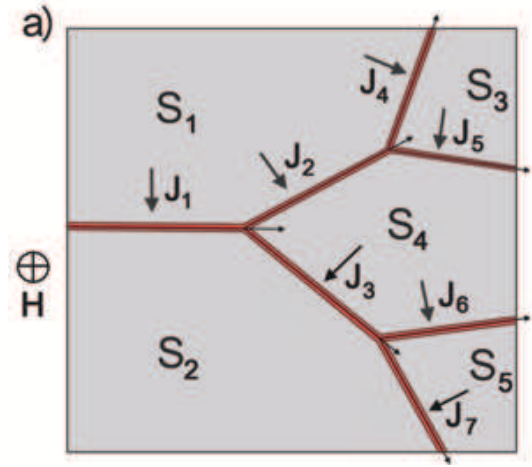}\\
\caption{(Color online) Tree-like branched Josephson junction.}
\label{tree}
\end{figure}

It is meaningful to compare the above results with those for their
linear (unbranched)counterpart considered in the Ref.\ci{Kupl07}.
Comparing dependence of $k_c$ on $L$ presented in Fig.\ref{pic4},
with the corresponding plot from for linear case,  one can
conclude that they are very close to each other. However,
differences between linear and branched cases appear in the plots
of $J_j(L)$ and $J_j(H)$ presented in Figs. \ref{pic2} -
\ref{pic7}, respectively. Comparing $J_j(H)$ in Figs. \ref{pic6}
and \ref{pic7} for branched Josephson junction with corresponding
plot in Fig.\ref{pic8} for linear case, one can find considerable
difference both in the shape and area of the stability region. In
particular, for branched case the total area of the stability
region is much larger than that for linear counterpart.  Moreover,
due to the fact that branched system has more parameters, one can
make it tunable with respect to playing with these parameters.
Especially, this concerns the case of more complicated branching
architecture, e.g., junction with tree-like branching presented in
Fig. \ref{tree}. Static solitons in this structure can be modeled
in terms of the
sine-Gordon equation with the boundary conditions given on metric tree graph.\\

\section{Conclusions} We have studied the current carrying states
in branched Josephson junction interacting with the external
magnetic field. The structure is assumed to be constructed,  from
three planar superconductors connected to each other via the
insulating (or normal metal) Y-junction. The system is modeled in
terms of the stationary sine-Gordon equation on the metric star
graph, whose solutions describe the phase difference between the
superconductors on the each branch of the junction. The boundary
conditions for the sine-Gordon equation at the branching point are
derived from the relation between current, local and external
magnetic fields.  Exact analytical solutions of sine-Gordon
equation fulfilling such boundary conditions are obtained. The
stability regions for these solutions are determined in terms of
the integration constant using the  Gibbs free energy functional
based (variational) approach. Physical observable values of the
current described in terms of the stable solutions are derived
explicitly as a function of the magnetic field. Finally, we note
that although we considered very simple branching having the form
of Y-junction, the approach we  used can be directly extended for
modeling static solitons in more general branching architectures
of the junction, such as tree, loop, triangle, etc. This can be
done similarly to that in \ci{Karim2018}, where sine-Gordon
equation on metric graphs is solved for $J_j=0$. Considering such
complicated branching architectures is of importance from the
viewpoint of the device tuning and
optimization in such problems as SQUID, superconducting qubit, cold atom trapping and Majorana wire networks.\\

%
%

%
%

\acknowledgments This work is supported by the grant  of the
Ministry for Innovation Development of Uzbekistan (Ref. No.
BF2-022).

\end{document}